\documentclass[letter,11pt]{article}
\pdfoutput=1

\usepackage{jheppub-mod}
\usepackage[T1]{fontenc}
\usepackage{amsthm}
\usepackage{listings}
\usepackage{mathrsfs}
\usepackage{xcolor}
\usepackage{color}
\usepackage{graphicx}
\usepackage{dcolumn}
\usepackage{bm}
\usepackage{physics}
\usepackage{amssymb,amsmath,amsfonts}
\usepackage{epsfig}
\usepackage{ccaption}

\def\be{\begin{equation}}
\def\ee{\end{equation}}
\def\ba{\begin{eqnarray}}
\def\ea{\end{eqnarray}}

\def\no{\nonumber \\}

 \def\ro{\rho}
 
 \def\a{{\alpha}}
 
 \def\frac#1#2{{#1\over #2}}

 \def\s{\sqrt}
 
 \def\b{{\beta}}
 \def\o{{\rm ord}}

 \def\p{\partial}

\begin{document}

\title{ Unitary Constraints on Semiclassical Schwarzschild Black Holes in the Presence of Island }

\author[a]{ Dong-Hui Du,}
\author[b]{ Wen-Cong Gan,}
\author[c,d,e,*]{ Fu-Wen Shu}
\author[a,*]{and Jia-Rui Sun}

\renewcommand{\thefootnote}{\fnsymbol{footnote}}
\footnotetext[1]{Corresponding authors.}
\renewcommand{\thefootnote}{\arabic{footnote}}

\affiliation[a]{School of Physics and Astronomy, Sun Yat-Sen University, Guangzhou, 510275, China}
\affiliation[b]{GCAP-CASPER, Physics Department, Baylor University, Waco, Texas 76798-7316, USA}
\affiliation[c]{Department of Physics, Nanchang University, Nanchang, 330031, China}
\affiliation[d]{Center for Relativistic Astrophysics and High Energy Physics, Nanchang University, Nanchang, 330031, China}
\affiliation[e]{Center for Gravitation and Cosmology, Yangzhou University, Yangzhou, China}

\emailAdd{donghuiduchn@gmail.com}
\emailAdd{Wen-cong$\_$Gan1@baylor.edu}
\emailAdd{shufuwen@ncu.edu.cn}
\emailAdd{sunjiarui@mail.sysu.edu.cn}


\abstract{
We reconsider $D\geq4$ dimensional asymptotically flat eternal Schwarzschild black hole, and focus on the situation where the inner boundary of the radiation region is chosen to be near the horizon (i.e. $\beta \ll1$). The tension between the near-horizon condition and the short-distance approximation emerges in large dimensions in $[JHEP 06 (2020) 085]$. We remove this tension by introducing a more proper near horizon condition, thus the resulting island solution is well-behaved in any $D\geq4$ dimensional spacetime. Interestingly, a novel constraint is obtained in this situation as required by the existence of the island solution, which directly leads to the constraints on the size of the Schwarzschild black hole, the position of the inner boundary for the radiation region, or the value of $c\cdot\tilde{G}_{N}$ in any $D\geq4$ dimension. When considering the large $D$ limit, the constraint on the size of the Schwarzschild black hole obtained in this situation is in agreement with the result given in $[Phys.Rev.D 102 (2020) 2, 026016]$. We interpret these as the unitary constraints implied by the presence of island in semiclassical gravity.
}

\maketitle

\section{ Introduction }
The black hole information paradox \cite{Hawking:1976ra} has been plaguing physics community for many years since Hawking raised it, as it seemingly indicates the breakdown of unitarity in semiclassical description of gravity. The paradox can be reflected by the monotonically increasing behavior of the entanglement entropy of radiation in Hawking's calculations. While the unitarity requires that the entanglement entropy of radiation should follow the so-called Page curve \cite{Page:1993wv,Page:2013dx}. Remarkably, recent breakthrough has been made in solving the black hole information paradox via semiclassical gravitational calculations \cite{Penington:2019npb,Almheiri:2019psf,Almheiri:2019hni}, benefiting from the RT formula \cite{Ryu:2006bv,Hubeny:2007xt} and its quantum corrected generalizations \cite{Faulkner:2013ana,Engelhardt:2014gca}. The proposed island formula in \cite{Penington:2019npb,Almheiri:2019psf,Almheiri:2019hni} directly came from the quantum extremal surface prescription \cite{Engelhardt:2014gca}, and was further confirmed by the gravitational replica calculations \cite{Penington:2019kki,Almheiri:2019qdq}. See recent works on islands for black holes in various gravitational theories \cite{Almheiri:2019yqk,Chen:2019uhq,Almheiri:2019psy,Gautason:2020tmk,Hashimoto:2020cas,Anegawa:2020ezn,
Hartman:2020swn,Hollowood:2020cou,Alishahiha:2020qza,Bak:2020enw,Dong:2020uxp,Krishnan:2020fer,
Chen:2020jvn,Ling:2020laa,Matsuo:2020ypv,Goto:2020wnk,Caceres:2020jcn,Karananas:2020fwx,Wang:2021woy,
Kim:2021gzd,Lu:2021gmv,Yu:2021cgi,Arefeva:2021kfx,He:2021mst,Matsuo:2021mmi,Arefeva:2022guf,
Arefeva:2022cam,Gan:2022jay,Tian:2022pso,Yadav:2022fmo,Anand:2022mla,Ageev:2022hqc,Geng:2020qvw,Geng:2020fxl,
Geng:2021hlu,Saha:2021ohr,Ahn:2021chg,Krishnan:2020oun,Yu:2021rfg,Omidi:2021opl}.

As a brand-new concept, the island formalism can reproduce the Page curve and help preserve the unitarity in semiclassical gravity. One may wonder whether there are more implications that the island can bring for us. In the present paper, we reconsider the island solution in asymptotically flat eternal Schwarzschild black hole when the inner boundary of the radiation region is chosen to be near the horizon, i.e $\b\ll1$. We will show the general island solution in situation $\b\ll1$, including the solution satisfying $\a\sim\b$. The tension between the original near-horizon condition and the short-distance approximation emerges in large dimensions in Ref. \cite{Hashimoto:2020cas}, and we remove this tension by introducing a more proper near horizon condition, see the Appendix \ref{Appendix A}. We will show a constraint on the Schwarzschild black hole in the presence of the island solution in situation $\b\ll1$, which we interpret as the unitary constraint in semiclassical gravity. Similar results may be obtained for other types of black holes.

This paper is organized as follows. In Section \ref{sec-island-formula}, we review the entropy formulas in Ref. \cite{Hashimoto:2020cas} used to calculate the entanglement entropy in Schwarzschild black hole spacetime. In Section \ref{sec-island-sulution}, we reconsider $D\geq4$ dimensional asymptotically flat eternal Schwarzschild black hole and we focus on the situation in which the inner boundary of the radiation region $R$ is chosen to be near the horizon. The general island solution in this situation is studied. Moreover, a general constraint is found as required by the existence of the island solution in this situation. Then in Section \ref{sec-constraint}, some implications from this constraint are discussed. In Section \ref{Page-Scrambling time}, the Page time is obtained while some doubts about the estimation of the scrambling time are raised in this situation. The conclusion and discussion are in the Section \ref{Conclusion-discussion}.

\section{ Island formula in Schwarzschild black hole spacetime }\label{sec-island-formula}
The island formula has been widely and successfully used to find the island solution in black hole spacetime. This gives us the entanglement entropy of the radiation as
\ba\label{Island-formula}
S_{R}=\text{min}\left\{\text{ext}\left[\frac{\text{Area}(\p I)}{4\tilde{G}_N}+S^{\text{finite}}_{\text{matter}}(I\cup R)\right]\right\},
\ea
where the optimal $\p I$ is given by the quantum extremal surface prescription \cite{Engelhardt:2014gca}. And the area-like divergence \cite{LRJR,MSre} of the entanglement entropy of matter fields has been absorbed into the renormalized Newton's constant $\tilde{G}_N$ \cite{LSJU}.

To evaluate the finite part of the entanglement entropy of matter fields in $D$-dimensional Schwarzschild black hole spacetime, it is useful to utilize the following two formulas corresponding to two limits \cite{Hashimoto:2020cas}:
\begin{itemize}
\item[(1)] At large distance limit, by assuming the s-wave approximation, the finite part of the matter's entanglement entropy can be approximated by the mutual information of the two-dimensional massless fields as \cite{Hashimoto:2020cas}
\be \label{Sm1}
 I(A:B) = - \frac{c}{3} \log d(A,B),
\ee
where $c$ is the central charge and $d(A,B)$ is the distance between the boundaries of region $A$, $B$.
\item[(2)] For sufficiently small distance $L$, the mutual information is approximately given by \cite{Hashimoto:2020cas,HCMH1,HCMH2}
\be \label{Sm2}
I(A:B) = c \kappa_{D} \frac{\mbox{Area}}{L^2},
\ee
where $c$ is the central charge of the free massless matter fields and $\kappa_{D}$ is a dimensionally dependent constant. To guarantee validity of this formula in curved spacetime, $L$ should be sufficiently smaller than the length scale of the curvature denoted as $l_{R}$, i.e. $L\ll l_{R}$.
\end{itemize}

\section{ Island for asymptotically flat eternal Schwarzschild black hole }\label{sec-island-sulution}
In the following, we consider the asymptotically flat eternal Schwarzschild black hole. The metric for the Schwarzschild black hole in $D\equiv n+3$ dimensions is
\be
ds^2 = -f(r) dt^2 + \frac{dr^2}{f(r)} + r^2 d\Omega_{n+1}^2, \quad f(r) \equiv 1-\left(\frac{r_{h}}{r}\right)^{n},
\ee
where $d\Omega_{n+1}$ is the line element of unit-sphere $S^{n+1}$. And the temperature of the Schwarzschild black hole is $\ T_{h} = {1}/{\beta_{h}} = {n}/{(4 \pi r_{h})}$.

We will adopt the coordinate $ \rho={(r/r_{h})}^n $ as introduced in large dimension gravity \cite{RRK,RDK}, where the space-time dimension can also be regarded as a variable. The metric of the Schwarzschild black hole is rewritten as
\ba
ds^2&=&-\frac{\rho-1}{\rho}dt^2+ \frac{r_{h}^2 \rho^{\frac{2}{n}}}{n^2\ro(\rho-1)}d\rho^2            +r_{h}^{2}\rho^{\frac{2}{n}}d\Omega_{n+1}^2,
\ea
with event horizon at $\rho=1$. Note that we still work in the general $D$-dimensional cases and will \textit{not} take the near-horizon and large dimension limit $D\to\infty$ as in Refs. \cite{RRK,RDK}. The reason why we choose to work in the coordinate $ \rho$ will be clear later. Now defining the tortoise coordinate as
\ba\label{tortoise}
r_{*}(\rho) &\equiv& \int^{r} \frac{1}{f(r(\rho))}dr = \int^{\rho} \frac{r_{h} \rho^{\frac{1}{n}}}{n(\rho -1)}\ d \rho\no
&=& -\frac{r_{h}\rho^{(n+1)/n}}{n+1}{}_{2}{F}_1[\ 1,1+\frac{1}{n},2+\frac{1}{n},\rho\ ],
\ea
where ${}_{2}{F}_1$ is the hypergeometric function. The Kruskal coordinates on the right wedge of the Penrose diagram are
\begin{align}
&U = -2r_{h}e^{-n\frac{t-r_*(\rho)}{2r_{ h}}},\ V = 2r_{h}e^{n\frac{t+r_*(\rho)}{2r_{ h}}}.\ (\text{outside horizon})
\end{align}
Thus the metric can be converted into
\ba
ds^2&=&-\frac{dU dV}{W(\rho)^2}+r_{h}^{2}\rho^{\frac{2}{n}}d\Omega_{n+1}^2,
\ea
where
\ba
W(\rho)\equiv n\sqrt{\frac{\rho}{\rho-1}}e^{n\frac{r_*(\rho)}{2r_{h}}}.
\ea

\subsection{ Without island  }
\begin{figure}
  \centering
  \includegraphics[width=0.6\textwidth]{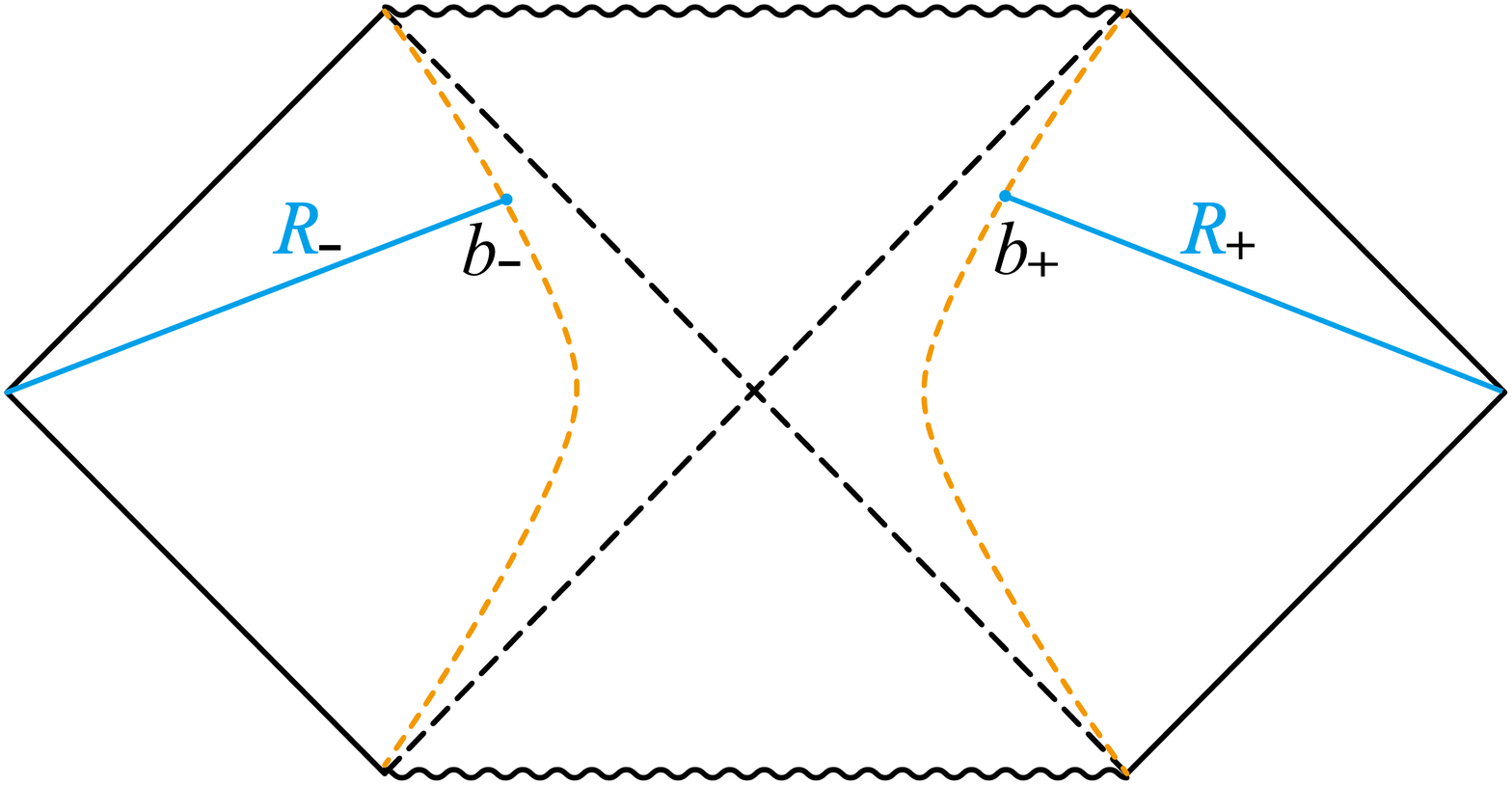}
  \caption{The Penrose diagram of the asymptotically flat eternal Schwarzschild spacetime
	without island. Where $b_{\pm}$ represent the inner boundaries of the radiation region $R=R_{+}\cup R_{-}$.}
\label{no-island}
\end{figure}

For the case without island, the total entanglement entropy of the radiation is purely given by the finite part entanglement entropy of the matter fields. We consider two inner boundary points of radiation's region $R=R_{+}\cup R_{-}$ on the right and left wedges of the Penrose diagram, denoted as $b_+=(t_b,r_b)$ and $b_-=(-t_b + i {\beta_{h}}/{2},r_b)$ respectively (see the Fig. \ref{no-island}). From \cite{Hashimoto:2020cas}, the total entanglement entropy of the radiation is
\ba
 S_R = S^{\text{finite}}_{\text{matter}}(R) = -I(R_{+}:R_{-}),
\ea
then by using the formula \eqref{Sm1} and working in the Kruskal coordinates, we obtain
\ba \label{S-no island}
 S_R &=& \frac{c}{6} \log \frac{\left[U(b_-) - U(b_+)\right]\left[V(b_+) - V(b_-)\right]}{W(b_+)W(b_-)}\no
   &=& \frac{c}{6} \log \left[\frac{16 r_{h}^2(\rho_b -1)\cosh^2\frac{n t_b}{2r_h}}{n^2\rho_b}\right].
\ea
At late time $t_{b}\gg r_{b}>r_{h}$, $S_R$ becomes
\ba \label{EE-no island}
S_R\approx  \frac{n c}{6r_{h}}t_{b},
\ea
for finite $r_{h}$ and $\rho_b>1$ in any dimension $n\geq1$ (as for the large dimension limit, one can omit the logarithm divergent term of $n$ relative to its linearly divergent term). The result leads to Hawking's result which violates the unitarity at late time.

\subsection{ With island }\label{With-island}

\begin{figure}
  \centering
  \includegraphics[width=0.6\textwidth]{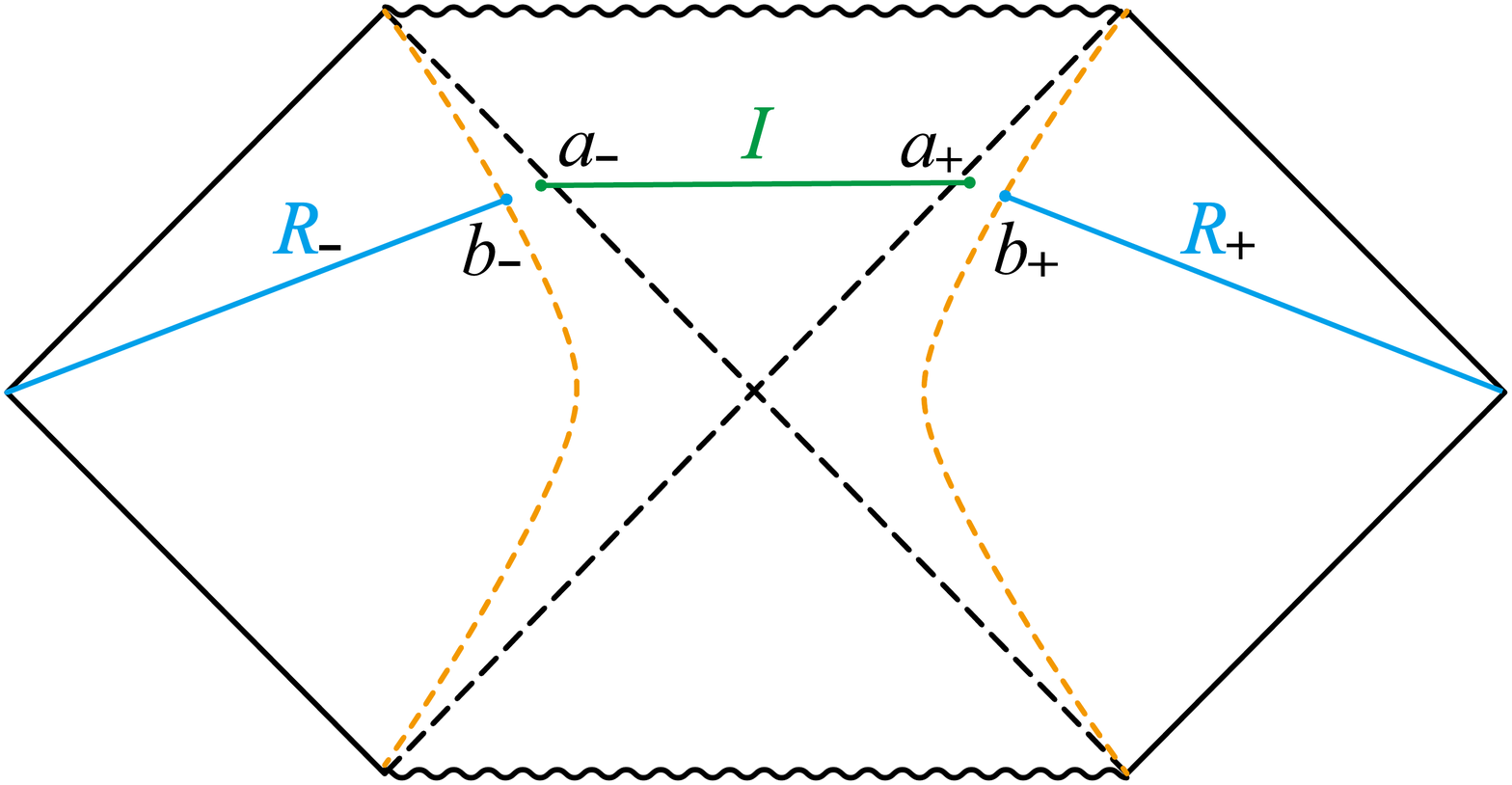}
  \caption{The Penrose diagram of the asymptotically flat eternal Schwarzschild spacetime
	with an island. Where $a_{\pm}$ and $b_{\pm}$ represent the boundaries of the island region $I$ and the inner boundaries of the radiation region $R=R_{+}\cup R_{-}$ respectively.  }
\label{with-island}
\end{figure}

We denote the boundary points of the island $I$ on the right and left wedges of the Penrose diagram as $a_+=(t_a,r_a)$ and $a_-=(-t_a + i {\b_{h}}/{2},r_a)$ respectively (see the Fig. \ref{with-island}). Here we assume the boundary of the island is outside and near the horizon. As elucidated detailedly in Appendix \ref{Appendix A}, to remove the tension between the original near-horizon condition used in Ref. \cite{Hashimoto:2020cas} and the short-distance approximation emerges in large dimensions, we will take $\a< \b \ll 1$ as the new near-horizon condition with parameters $\a\equiv\s{\ro_a-1}>0$ and $\b\equiv\s{\ro_b-1}>0$, where $\ro_a={(r_{a}/r_{h})}^n$ and $\ro_b={(r_{b}/r_{h})}^n$. So that the geodesic distance can be estimated well by the first order term of $\a, \b$, i.e.
\ba  \label{Lab3}
L &\simeq& \frac{2 r_{h}(\b-\a)}{n},
\ea
which is valid for any $n \geq 1$ even for large dimension case. Remember that as required by the validity of the formula \eqref{Sm2}, we must ensure $L\ll l_{R} $. Note that the characteristic curvature length outside and near the horizon is \cite{RRK}
\ba \label{LR}
l_{R}\simeq K^{-1/4}=\frac{r_{h}}{\sqrt[4]{n(n+1)^2(n+2)}}\sim\frac{r_{h}}{n},
\ea
where $K\equiv R_{\mu\nu\rho\sigma}R^{\mu\nu\rho\sigma}$ is the Kretschmann scalar. By comparing with formulas \eqref{Lab3} and \eqref{LR}, one can find the condition $(\b-\a)< \b \ll 1$ is also sufficient to guarantee $L\ll l_{R} $ even in large dimensions.

With the island, the finite part of the matter's entanglement entropy is given by \cite{Hashimoto:2020cas}
\ba  \label{Sm22}
S^{\text{finite}}_{\text{matter}}(R\cup I) = -2I(R_{+}:I).
\ea
Then by the formulas \eqref{Sm2} and \eqref{Sm22}, the total entanglement entropy of the radiation in coordinate $\ro$ is equal to
\ba  \label{S-island}
S_R&=&\frac{\Omega_{n+1}r^{n+1}_{h}\ro_a^{\frac{n+1}{n}}}{2l_{p}^{n+1}}-2 c \kappa_{D}\frac{\Omega_{n+1}r^{n+1}_{h}\ro_b^{\frac{n+1}{n}} }{L^{n+1}},
\ea
where $\tilde{G}_{N}\equiv l_{p}^{n+1}$ with the Planck length $l_{p}$, and the factor 2 is due to the double contributions from the left and right wedges. Here $\Omega_{n+1}\equiv {2\pi^{(n+2)/2}}/{\Gamma[(n+2)/2]}$ is the volume of unit-sphere $S^{n+1}$.

To find the island solution, we put $\ro_a=1+\a^2,\ \ro_b=1+\b^2$ and \eqref{Lab3} into the formula \eqref{S-island}, we have
\ba  \label{Sab}
S_R \simeq \frac{\Omega_{n+1}}{2}\left[\left(\frac{r_{h}}{l_{p}}\right)^{n+1}
-\frac{c \kappa_{D} n^{n+1}} {2^{n-1}} {(\b-\a)^{-(n+1)} }\right],
\ea
where we used the condition $\a\ll 1$ and $\b\ll 1$. Then in order to find the minimal value of $S_R$, we take the derivative with respect to $\a$ and solve the following equation, as
\ba \label{pSab}
\p_{\a} S_R = \frac{(n+1)\Omega_{n+1}\b}{n}\left[ \frac{\a}{\b} \left(\frac{r_{h}}{l_{p}}\right)^{n+1} -\frac{c \kappa_{D} n^{n+2}}{2^{n}\b^{n+3}} {(1-\frac{\a}{\b})^{-(n+2)} }\right]
&=& 0.
\ea
However, it is hard to give an exact analytical solution due to the complex form of above equation for any $n\geq1$. We further assume the condition $\a \ll \b $ \footnote{ In fact, the island solution in Ref. \cite{Hashimoto:2020cas} was similarly obtained under the assumption $\tilde{\a} \ll \tilde{\b}$ (with $\tilde{\a} \equiv \sqrt{{(r_{a} -r_{h})}/{r_h}}$ and $\tilde{\b} \equiv \sqrt{{(r_{b} -r_{h})}/{r_{h}}}$) by implicitly assuming that $r_{b}-r_{h}$
is much larger than the Planck length, as $r_{a}-r_{h}\sim {\cal O}(\tilde{G}^{2}_{N}) \ll  r_{b}-r_{h} \sim {\cal O}(l^{0}_{p})$. As we will discuss later, it is necessary to investigate the general island solution for $\tilde{\a} < \tilde{\b}$. While we use $\a, \b$ in the paper. } and expand the Eq. \eqref{pSab} to the first order of small $\a/\b$, which is equivalent to
\ba \label{pSab-n}
\frac{\a}{\b} \left(\frac{r_{h}}{l_{p}}\right)^{n+1} -\frac{c \kappa_{D} n^{n+2}}{2^{n}\b^{n+3}} \left[1+(n+2)\frac{\a}{\b}+ \mathcal{O}\left(\left(\frac{n\a}{\b}\right)^2\right)\right]
= 0,
\ea
where we used $n\gg1$ in order to see the large dimension behavior. A more specific condition $\a \ll \b/n $ is required so that we can omit the higher order terms even for large dimension case. We obtain an island solution as
\ba \label{qes}
\a &\simeq& \frac{\b}{\frac{2^{n}}{c\kappa_{D}n^{n+2}}\left(\frac{r_{h}}{l_{p}} \right)^{n+1}\b^{n+3}-(n+2)},
\ea
which is valid under the condition $\a \ll \b/n \ll 1$ and also can be applicable for the large dimension case. \footnote{It is worth mentioning that there also exists an island solution in large dimension case $n\rightarrow\infty$ if assuming $\a\sim \b/n$ (differing from the solution \eqref{qes} under assumption $\a\ll \b/n$ for any $n\geq1$). In this case, we can assume $ \frac{n\a}{\b}=\lambda\sim{\cal O}(1)$, thus $(1-\frac{\a}{\b})^{-(n+2)}=(1-\frac{\lambda}{n})^{-(n+2)}\simeq e^{\lambda}$ in the limit $n\rightarrow\infty$. Then by solving Eq. \eqref{pSab}, two branches of real number solutions can be found, i.e. $\a_{1} \simeq -\frac{\b}{n}\text{Productlog}\left[-\frac{1}{X}\right]$ and $\a_{2} \simeq -\frac{\b}{n}\text{Productlog}\left[-1, -\frac{1}{X}\right]$, where $X$ is a real value defined in \eqref{X0}. In this case, $S_{R}$ reaches its local minimum and local maximum at $\a_{1}$ and $\a_{2}$ respectively, thus the island solution is given by $\a_{1}$ (which are valid for $n\rightarrow\infty$ while $\a\sim \b/n$). Meanwhile the existence of this real number solution requires the constraint $X > e$, which is in keeping with the constraint \eqref{bound-Xn} as $X_{c}\rightarrow e$ for $n\rightarrow\infty$.} This result only differs from the result of Ref. \cite{Hashimoto:2020cas} by a next-to-leading order correction in finite lower dimensions. With this solution, one may assume the condition $c\tilde{G}_N/{r_{h}^{n+1}}={c l_{p}^{n+1}}/{r_{h}^{n+1}}\ll1$ to make semiclassical approximation and drop the second term in the denominator of \eqref{qes}. However, as $\b$ itself can be very small, the condition ${c l_{p}^{n+1}}/{r_{h}^{n+1}}\ll1$ is not a sufficient condition to drop the second term in the denominator. And we do not make such an approximation in the paper. Furthermore, by putting the condition $\a \ll \b/n $ on the result \eqref{qes} to make sure the consistency, we obtain that
\ba \label{bound0}
\b^{n+3}\left(\frac{r_{h}}{l_{p}} \right)^{n+1} \gg \frac{c\kappa_{D}(n+1)n^{n+2}}{2^{n-1}}.
\ea
The constraint \eqref{bound0} is required by the existence of the consistent island solution for the special case $\a \ll \b/n\ll1$. By plugging the solution \eqref{qes} back into the formula \eqref{Sab}, the entanglement entropy in this special case is given by
\ba \label{EEa}
S_R &\simeq& \frac{\Omega_{n+1}}{2}\left(\frac{r_{h}}{l_{p}}\right)^{n+1}
\left[1- \frac{2\b^2}{n(n+3)}\frac{c\kappa_{D}(n+3)n^{n+2}}{2^n\b^{n+3}\left(\frac{r_{h}}{l_{p}} \right)^{n+1}}\right. \no
&& \left.\times \frac{\left(2^n\b^{n+3}\left(\frac{r_{h}}{l_{p}} \right)^{n+1}-c\kappa_{D}n^{n+2}\right)}{\left(2^n\b^{n+3}\left(\frac{r_{h}}{l_{p}} \right)^{n+1}-c\kappa_{D}(n+2)n^{n+2}\right)}\right].
\ea

Note that the island solution \eqref{qes} and the constraint \eqref{bound0} are only valid for the special case $\a \ll \b/n\ll1$. However, we wonder whether there exists a general island solution and a general constraint for $\a < \b\ll1$. Because it is reasonable to take $\b$ to be small enough so that $\b$ get close to $\a$, i.e. $\a \sim \b$. This requires us to solve Eq. \eqref{pSab} without assuming $\a \ll \b/n$, which is poorly investigated in higher dimensions in previous papers but has been studied in $D=4$ case in Ref. \cite{Matsuo:2020ypv}. Motivated by this consideration, we reanalyze Eq. \eqref{pSab} with $\a<\b \ll 1$, which can be equivalently written as
\ba \label{pSab-x}
x(1-x)^{n+2} = \frac{c \kappa_{D} n^{n+2}}{2^{n}\b^{n+3} \left(\frac{r_{h}}{l_{p}}\right)^{n+1}}= \frac{1}{(n+3)X},
\ea
where we define $x \equiv \a/\b \in (0, 1)$ and
\ba \label{X0}
X = X(n,c,\b,\frac{r_{h}}{l_{p}}) \equiv \frac{2^{n}\b^{n+3} \left(\frac{r_{h}}{l_{p}}\right)^{n+1}}{c \kappa_{D}(n+3) n^{n+2}}.
\ea

\begin{figure}
    \begin{tabular}{cc}
\includegraphics[height=5.2cm]{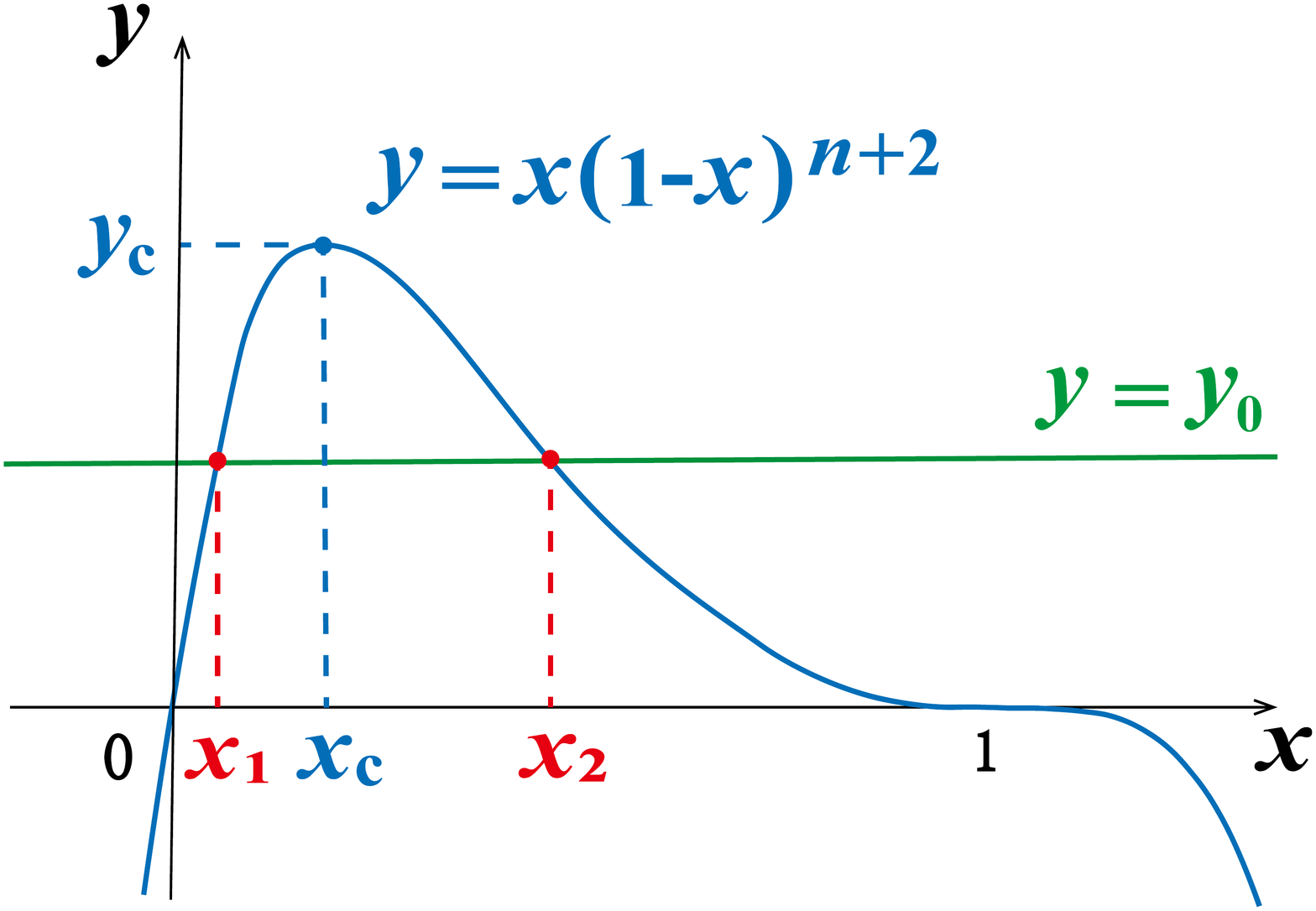}&
\includegraphics[height=5.2cm]{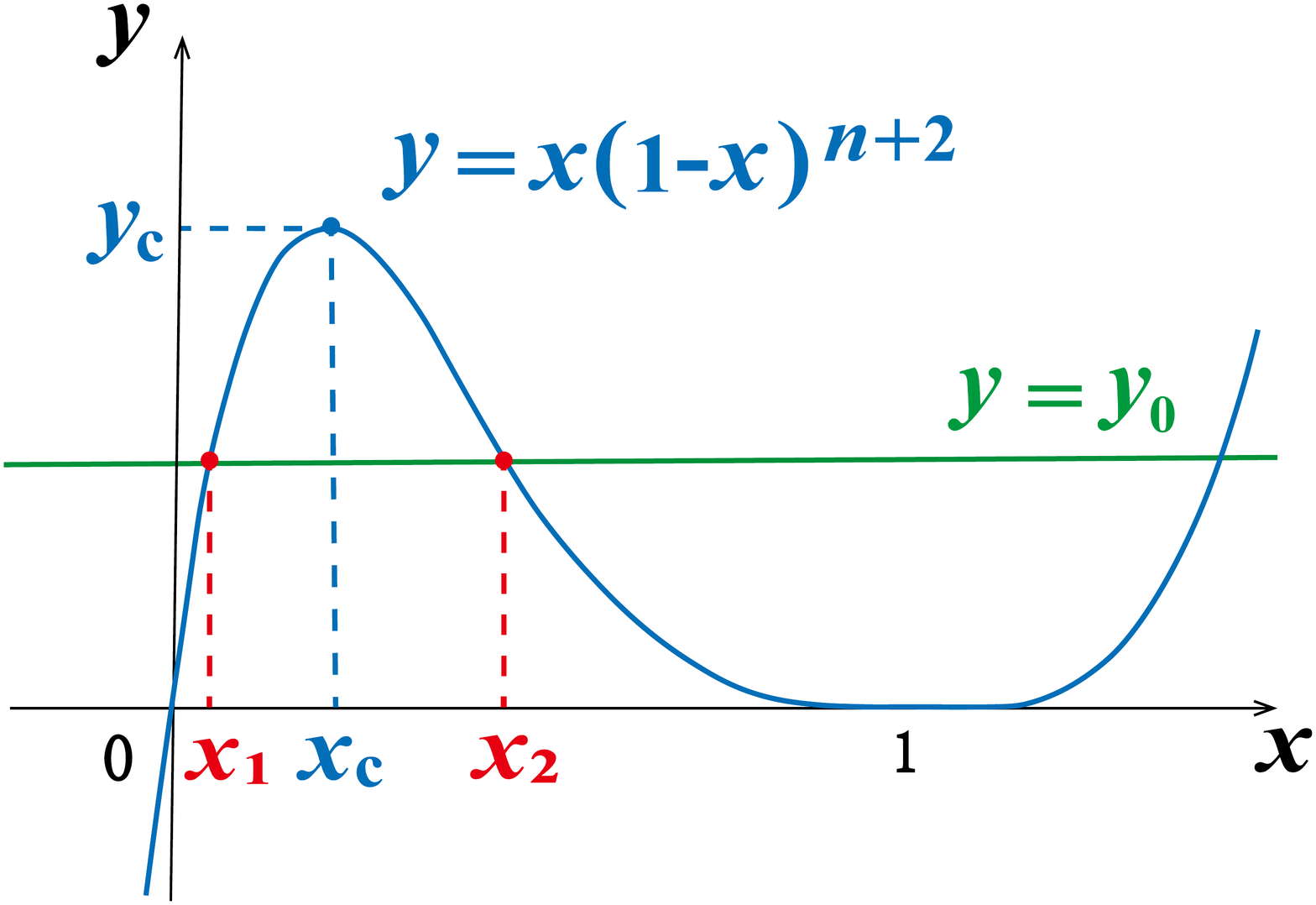}\\
	(a) For odd $n$. & (b) For even $n$.  \\[6pt]
	\end{tabular}
\caption{Island solution for general case $\a<\b\ll1$, where $y_{0}=1/[(n+3)X].$ }
\label{island-solution}
\end{figure}

The existence of the island solution for general $\a<\b\ll1$ can be easily confirmed (see the Fig. \ref{island-solution}). In the interval $x \in (0, 1)$, one can find the function $y= x(1-x)^{n+2}$ is monotonically increasing with $x$ in the interval $(0, x_{c})$ and monotonically decreasing with $x$ in the interval $(x_{c},1)$. A local maximum of function $y$ can be found as
\ba
y_{c}=\frac{(n+2)^{n+2}}{(n+3)^{n+3}},\ \text{at}\ x_{c}= \frac{1}{n+3}.
\ea
And the constraint $y_{0} < y_{c}$ is required to make sure the existence of the island solution for this situation, i.e.
\ba \label{bound-Xn}
X > \left(1+\frac{1}{n+2}\right)^{n+2} \equiv X_{c}.
\ea
Where $X_{c}$ has a finite range as $64/27\leq X_{c}< e$ for any $n\geq1$. The constraint \eqref{bound0} obtained by assuming $\a\ll \b/n$ only reflects the limited behavior of the constraint \eqref{bound-Xn}, as
\ba \label{bound-limt}
X \gg X_{c}>\frac{2(n+1)}{n+3}\geq1,
\ea
for any $n\geq1$. And the result \eqref{qes} only represents a special class of solutions subject to the condition $X \gg X_{c}>1$, i.e. $y_{0}\rightarrow 0$ (see the Fig. \ref{island-solution}). We will discuss the constraint \eqref{bound-Xn} in more details in next section.

\begin{figure}
    \begin{tabular}{cc}
\includegraphics[height=5.3cm]{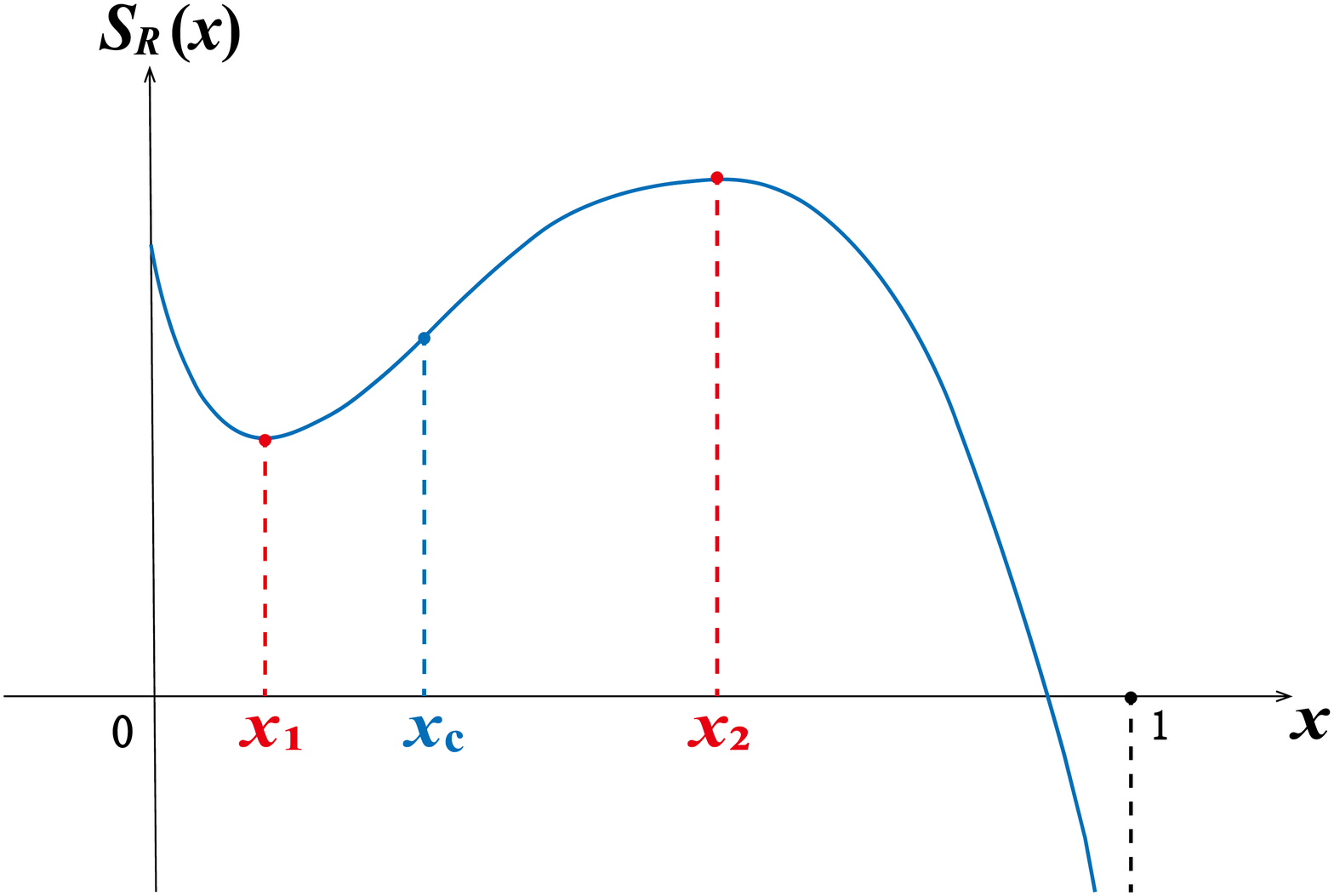}&
\includegraphics[height=5.3cm]{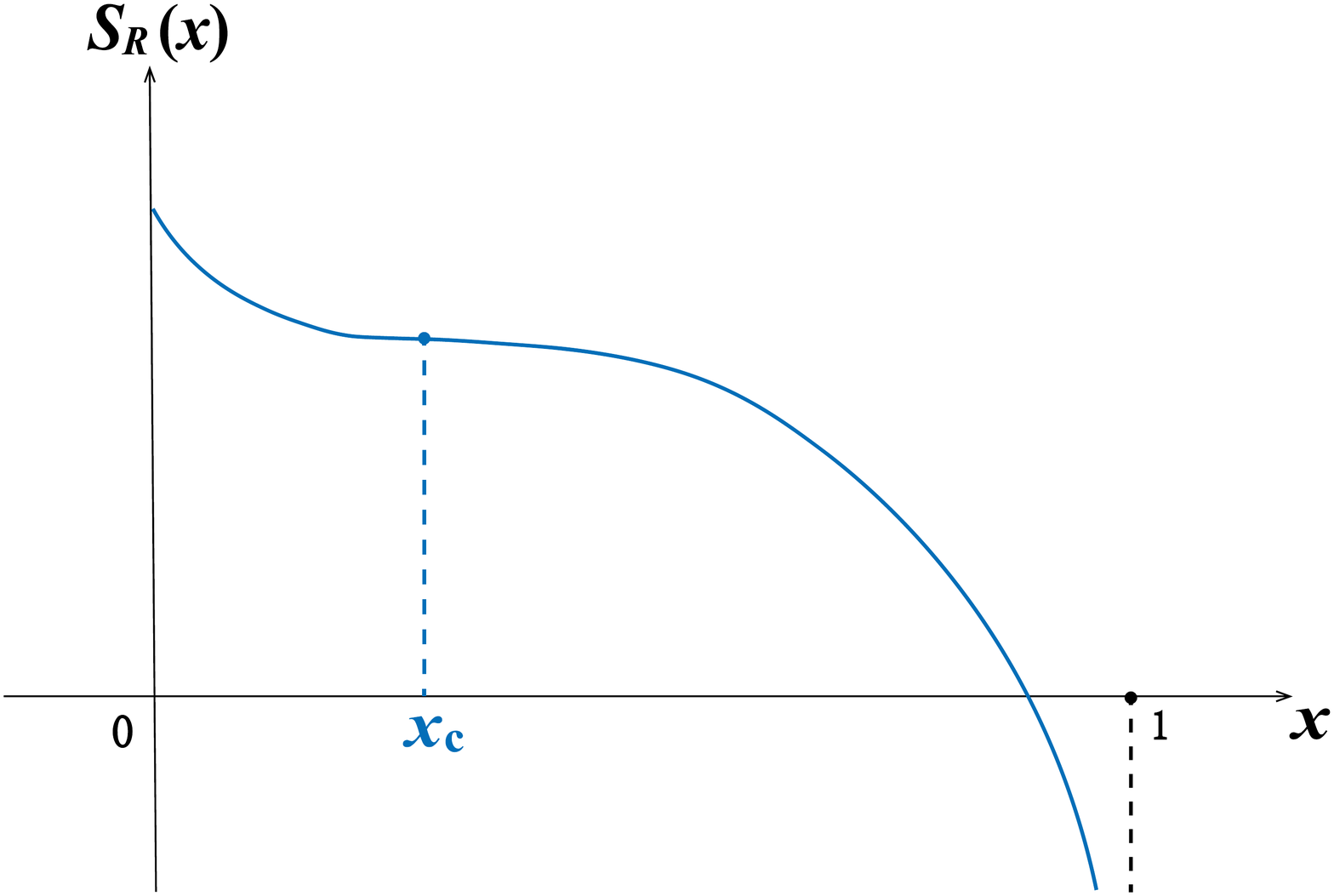}\\
	(a) $X>X_{c}$. & (b) $X \leq X_{c}$.  \\[6pt]
	\end{tabular}
\caption{ The entropy curve of $S_{R}$ with respect to $x =\a/\b$ for $X>X_{c}$ and $X\leq X_{c}$. Where $S_{R}\rightarrow -\infty$ for $x\rightarrow1$ (i.e. $\a\rightarrow\b$) by formula \eqref{Sab}.  }
\label{entropy}
\end{figure}

Once the constraint \eqref{bound-Xn} is satisfied, two solutions $x_{1}$ and $x_{2}$ (where $x_{1}<x_{c}<x_{2}$) can be found for $x\in(0, 1)$. The entanglement entropy $S_{R}$ reaches its local minimum and local maximum at $x_{1}$ and $x_{2}$ respectively. The island solution is exactly given by $x = x_{1}$, whose exact value is determined by the value of $y_{0}$. Therefore, we confirm the existence of the island solution for general case $\a<\b \ll 1$ satisfying $X>X_{c}$, where we have
\ba \label{general-solution}
0< \frac{\a}{\b}< x_{c}=\frac{1}{n+3}\ \text{and}\ \frac{\a}{\b}=
\begin{cases}
\frac{1}{(n+3)X-(n+2)},\ \text{for}\ X\gg X_{c} \\
\frac{1}{n+3},\ \text{for}\ X\rightarrow X_{c}
\end{cases}
\ea
by considering the solution \eqref{qes}. However, it is hard to give a general analytical expression of the island solution for $\a<\b \ll1$. Note if we take $X = X_{c}$, $x_{1}$ and $x_{2}$ will meet at $x=x_{c}$, then there would be no local minimum for $S_{R}$ at $x_{1}$, and $S_{R}$ will monotonically decrease with $x$ for $x\in(0, 1)$, thus there is no nontrivial island solution. And $S_{R}$ will also monotonically decrease with $x$ if $X < X_{c}$ (see the Fig. \ref{entropy}), without nontrivial island solution. This analysis is essentially the same as in Ref. \cite{Matsuo:2020ypv} for $D=4$ case, here we generalize it to higher dimensions.

\section{ Unitary constraints in the presence of island }\label{sec-constraint}

The constraint \eqref{bound-Xn} is the significant point in the paper, as it can be understood as an unitary constraint from the presence of the island in semiclassical gravity. The constraint \eqref{bound-Xn} is equal to
\ba \label{novel-bound}
\b^{n+3}\left(\frac{r_{h}}{l_{p}} \right)^{n+1} &>& \frac{c\kappa_{D}(n+3)n^{n+2}X_{c}}{2^n}\equiv c \cdot \chi,
\ea
where $\chi\equiv {\kappa_{D}(n+3)n^{n+2}X_{c}}/{2^n} $, which is dependent on the spacetime dimension (see the value of dimensionally dependent constant $\kappa_{D}$ in Ref. \cite{HCMH2}). Remember this result is only valid for the situation $\b\ll 1$, and our following discussions are limited to this situation. There are some aspects from the constraint \eqref{novel-bound} in this situation as shown in the following:

\begin{itemize}
\item  Constraint on $r_{h}$ when $n,\ c$ and $\b$ are fixed

From the constraint \eqref{novel-bound}, for fixed $n,\ c$ and $\b$, we can obtain
\ba \label{bound1}
\frac{r_{h}}{l_{p}}&>& \frac{(c \cdot\chi)^{\frac{1}{n+1}}}{\b^{\frac{n+3}{n+1}}}
\gg (c \cdot\chi)^{\frac{1}{n+1}},
\ea
which is valid for $\b \ll1$. We substituted the condition $\b\ll1$ into the inequality in order to show a general result in this situation, i.e. ${r_{h}}/{l_{p}}\gg (c \cdot\chi)^{\frac{1}{n+1}}$, and we allow the value of $r_{h}/{l_{p}}$ can be varied. The above constraint can not hold for the black hole whose size satisfying ${r_{h}}/{l_{p}} \sim (c \cdot\chi)^{\frac{1}{n+1}}$. For such black holes, there would be no consistent island solution and thus no well-behaved Page curve for the situation $\b\ll 1$. The constraint ${r_{h}}/{l_{p}}\gg (c \cdot\chi)^{\frac{1}{n+1}}$ can be regard as a physical constraint on the size of the black hole implied by the existence of the island in situation $\b\ll 1$, in order to recover unitarity in this situation. One may wonder whether $(c \cdot\chi)^{\frac{1}{n+1}}$ represents a strict and universal lower bound on the value of ${r_{h}}/{l_{p}}$ for any $\b>0$, however, it is not clear so far. As it requires to concretely calculate the entanglement entropy and give the complete constraint by the existence of the island solution for any $\b>0$, which would be complicated. In Refs. \cite{Arefeva:2021kfx,Arefeva:2022guf,Arefeva:2022cam,Ageev:2022hqc}, an explosive behavior of the entropy curve for small black holes (with $r_{h}\rightarrow 0$) in $D=4$ dimensions has been noticed, when the inner boundary of radiation region was far from the horizon, i.e. $\b\gg1$. These prompt us to investigate whether there is a strict and universal lower bound on the size of the black hole implied by the existence of the island for any $\b>0$.

\item  Constraint on $r_{h}$ when $c$ and $\b$ are fixed but $n\rightarrow \infty$

We note that the value of $(c \cdot\chi)^{\frac{1}{n+1}}$ in \eqref{bound1} will increase with the spacetime dimension. If we consider the large dimension case and utilize the approximation $\kappa_{D}=\Gamma[\frac{n+1}{2}]/(2^{n+5}\pi^{(n+1)/2})$ \cite{HCMH2} in large dimensions, where $D=d+1=n+3$. For finite $c$ and $n\rightarrow \infty$, we find
\ba \label{large-bound1}
\frac{r_{h}}{l_{p}} \gg (c \cdot\chi)^{\frac{1}{n}} \rightarrow \frac{n^{{3}/{2}}}{\sqrt{32\pi e}}.
\ea
This constraint for large dimension black holes is exactly in agreement with the constraint found in Ref. \cite{FDN} as $\frac{r_{h}}{l_{p}} \gtrsim n^{{3}/{2}}$, which was mainly motivated by the requirement of unitarity that the scrambling time should not exceed the evaporating time of the semiclassical black holes in large dimensions. As we discussed before, the existence of the island in situation $\b\ll 1$ provides the constraint ${r_{h}}/{l_{p}}\gg (c \cdot\chi)^{\frac{1}{n+1}}$, but it does not mean $(c \cdot\chi)^{\frac{1}{n+1}}$ is a strict and universal lower bound on ${r_{h}}/{l_{p}}$ for any $\b>0$. It would be interesting to further investigate why our constraint gives the consistent bound as in Ref. \cite{FDN} for large dimension black holes. Which may help figure out whether $(c \cdot\chi)^{\frac{1}{n+1}}{l_{p}}$ represents a strict and universal lower bound on the size of black holes for any $\b>0$.

Note that our result \eqref{large-bound1} is much tighter than that of Ref. \cite{FDN}. A possible interpretation is that the black hole with the size ${r_{h}}/{l_{p}} \sim (c \cdot\chi)^{\frac{1}{n+1}}$ in \eqref{bound1} has the manifest quantum gravity effects, where the near-horizon semiclassical approximation is invalid. As it will be clear in the last point that the constraint \eqref{novel-bound} can indeed be regarded as a semiclassical constraint to make sure the validity of semiclassical approximation for situation $\b\ll1$. It indicates that the quantum-gravity effects of the large-dimension black holes would be obvious at a much larger scale than we usually expect in low dimension case, as $r_{h}\sim n^{{3}/{2}}l_{p}\gg l_{p}$ when $n\rightarrow\infty$.

\item  Constraint on $\b$ when $n,\ c$ and $r_{h}$ are fixed

From the constraint \eqref{novel-bound}, for fixed $n,\ c$ and $r_{h}$, we can obtain
\ba \label{bound2}
\b &>& (c \cdot\chi)^{\frac{1}{n+3}}{\left(\frac{l_{p}}{r_{h}}\right)^{\frac{n+1}{n+3}}}\equiv \b_{\text{min}},
\ea
which can be transformed into $r_{b}$ by the definition $\rho_b={(r_b/r_{h})}^n =1+\b^2$, i.e.
\ba \label{bound2-r}
r_b &>&  \left(\left(\frac{c\kappa_{D}(n+3)^{n+3}n^{n+2}l_{p}^{n+1} }{2^n(n+2)^{n+2}r_{h}^{n+1}} \right)^{\frac{2}{n+3}}+1\right)^{\frac{1}{n}}r_{h}
= (r_b)_{\text{min}}.
\ea
This puts a lower bound on the position of the inner boundary of the selected radiation region. Note $\a /\b$ will increase monotonically when $\b$ is decreasing to $\b_{\text{min}}$ (as $y_{0}$ is increasing), until $\a/\b \rightarrow 1/(n+3)$ (as shown in the Fig. \ref{island-solution}). For $D=4$ dimension, if $\b=\b_{\text{min}}$, we will obtain
\ba
r_a = r_{h} + \sqrt{\frac{c\kappa_{4}}{54}}l_{p}\ \text{and}\ r_b=(r_b)_{\text{min}} = r_{h} + \sqrt{\frac{128 c\kappa_{4}}{27}}l_{p},
\ea
where the proper distance between the horizon and the inner boundary of the radiation region is of order $\mathcal{O}\left(\sqrt{r_{h}l_{p}}\right)$ (which is much larger than the Planck scale for the black hole satisfying $r_{h}/l_{p}\gg \sqrt{128c\kappa_{4}/27}\sim \mathcal{O}(1)$ from \eqref{bound1}). We conclude that $(r_b)_{\text{min}}$ is an exact solution of Eq. \eqref{pSab-x}. Note that in Ref. \cite{Matsuo:2020ypv}, a critical value $b_c$ (i.e. Eq. (4.3) in Ref. \cite{Matsuo:2020ypv}) was also found for $r_b$ in $D=4$ dimension by solving the same equation as Eq. \eqref{pSab-x} (i.e. Eq. (4.2) in Ref. \cite{Matsuo:2020ypv}). Nevertheless, one can check that $b_c$ is not an exact solution of Eq. \eqref{pSab-x} for $D=4$ (by substituting Eq. (4.3) to Eq. (4.2) in Ref. \cite{Matsuo:2020ypv}). Thus $(r_b)_{\text{min}}$ in Eq. \eqref{bound2-r} takes over the role of $b_{c}$ and related discussion of Ref. \cite{Matsuo:2020ypv} on the stretched horizon can also be drawn in higher dimensions.

\item  Constraint on $c$ (or $c \tilde{G}_{N}$) when $n,\ r_{h}$ and $\b$ are fixed

Note that we do not assume the semiclassical approximation ${cl_{p}^{n+1}}/{r_{h}^{n+1}}\ll1$ that was usually assumed when solving the equation. Instead, we can infer this condition from the constraint \eqref{novel-bound} for the situation $\b\ll1$. For fixed $n,\ r_{h}$ and $\b$, we can obtain
\ba \label{Semibound}
\frac{c l_{p}^{n+1}}{r_{h}^{n+1}}<\frac{2^{n}\b^{n+3}}
{\kappa_{D}(n+3)n^{n+2}X_{c}}\ll\frac{2^{n}}{\kappa_{D}(n+3)n^{n+2}X_{c}}=\frac{1}{\chi}.
\ea
Again we substituted the condition $\b\ll1$ into the inequality in order to show a general result in this situation, i.e. ${cl_{p}^{n+1}}/{r_{h}^{n+1}}\ll1/\chi$. Let us see the first inequality in \eqref{Semibound}, it will be tighter than the condition ${cl_{p}^{n+1}}/{r_{h}^{n+1}}\ll1$ due to the extremely tiny factor $\b^{n+3}$ for finite $n$. Or namely, profiting by the constraint \eqref{Semibound}, a sufficiently small $\b\ll1$ will ensure the semiclassical approximation ${cl_{p}^{n+1}}/{r_{h}^{n+1}}\ll1$. This indicates that the constraint \eqref{novel-bound} can be regarded as a semiclassical constraint for situation $\b\ll1$. Moreover, the constraint \eqref{Semibound} becomes tighter in higher dimensions as $1/\chi$ will decrease sharply when $n$ is increasing, and ${cl_{p}^{n+1}}/{r_{h}^{n+1}}\ll1/\chi\ll1$ for $n\rightarrow\infty$. It is interesting that this large dimension behavior will be obviously different from our usual expectation ${cl_{p}^{n+1}}/{r_{h}^{n+1}}\ll1$ in lower dimensions.

The constraint ${cl_{p}^{n+1}}/{r_{h}^{n+1}}\ll1/\chi$ on the value of the ${cl_{p}^{n+1}}$ (with fixed $r_{h}$) is required by the existence of the island in situation $\b\ll 1$, where we allow the value of ${cl_{p}^{n+1}}$ can be varied. Once ${cl_{p}^{n+1}}/{r_{h}^{n+1}}\sim1/\chi$, there would be no consistent island solution can be found in this situation. In Ref. \cite{He:2021mst}, similar conclusion about the existence of the upper bound on ${cl_{p}^{n+1}}$ (or $c\tilde{G}_{N}$) was also drawn for the Schwarzschild black hole and numerically checked in two-dimensional generalized dilaton theories, when the inner boundary of the radiation region was far from the horizon, i.e. $\b\gg1$. Here we give the $\b$-dependent upper bound of ${cl_{p}^{n+1}}$ (i.e. the first inequality in \eqref{Semibound}) for situation $\b\ll1$, while it is still not clear whether ${r_{h}^{n+1}}/\chi$ represents a strict and universal upper bound on the value of ${cl_{p}^{n+1}}$ for any $\b>0$. Remember that this discussion about the constraint on ${cl_{p}^{n+1}}$ is equivalent to the discussion about the constraint on $r_h$ through Eq. \eqref{novel-bound} for situation $\b\ll1$, which encourages us to think about whether there is also a connection between the result of Refs. \cite{Arefeva:2021kfx,Arefeva:2022guf,Arefeva:2022cam,Ageev:2022hqc} and that of Ref. \cite{He:2021mst} for situation $\b\gg1$.

\end{itemize}

In summary, the constraint \eqref{bound-Xn} shows us some unitary implications on the size of Schwarzschild black hole, the position of the inner boundary for the selected radiation region, or the value of $c\tilde{G}_{N}$ for situation $\b\ll1$. Which indicates that the above three seemingly different aspects \eqref{bound1}, \eqref{bound2-r} and \eqref{Semibound} can be attributed to the same origin, i.e. the unitary constraint \eqref{bound-Xn} in semiclassical gravity. However it remains to study the situation beyond $\b\ll1$, which may help us reveal more aspects in the presence of island.

\section{ Page time and scrambling time }\label{Page-Scrambling time}
\label{sec-5}
Let us evaluate the Page time. For the special case $\a\ll\b/n\ll1$ with the existence of an island, the corresponding island solution \eqref{qes} can be written as
\ba\label{qes-X}
\a &\simeq& \frac{\b}{(n+3)X-(n+2)}\simeq \frac{\b}{(n+3)X},
\ea
where we used $X \gg 1$ from \eqref{bound-limt} to drop the second term in the denominator, and note that the condition ${cl_{p}^{n+1}}/{r_{h}^{n+1}}\ll1$ is not a sufficient condition to drop that term. For this special case, the corresponding entanglement entropy \eqref{EEa} can be estimated as
\ba \label{EEb}
S_R
&\simeq& 2S_{BH}\left(1-\frac{2\b^2}{n(n+3)}\frac{1}{X}\frac{(n+3)X-1}{(n+3)X-(n+2)}\right) \no
&\simeq& 2S_{BH}\left(1-\frac{2\b^2}{n(n+3)}\frac{1}{X}\right) \simeq 2S_{BH},
\ea
where we used $X\gg1$ and $\b/n\ll1$. Therefore, for the special case $\a\ll\b/n\ll1$, the Page curve can be reproduced and the Page time can be obtained by combining \eqref{EE-no island} and \eqref{EEb}, thus
\ba \label{pagetime}
t_{\text{Page}} \simeq \frac{3}{\pi c}\frac{S_{BH}}{T_{h}}.
\ea
Furthermore, we can show that the approximate result \eqref{pagetime} still holds for general case $\a < \b \ll 1$, even without general analytical expression of the island solution for this case. By considering the entropy formula \eqref{Sab} and using \eqref{X0}, we have
\ba \label{EEc}
S_R &=& 2S_{BH}\left( 1
-\frac{1} {X} \frac{2\b^{2}}{n(n+3)} {\frac{1}{(1-\frac{\a}{\b})^{n+1}}}\right) \no
&>& 2S_{BH}\left( 1
-\frac{X_{c}} {X} \frac{2\b^{2}}{n(n+3)} {\frac{n+2}{n+3}}\right) \no
&>& 2S_{BH}\left( 1
-\frac{X_{c}} {X} \frac{2\b^{2}}{n(n+3)} \right)\simeq 2S_{BH}.
\ea
where in the first inequality we utilized the fact $0<\a/\b<1/(n+3)$ from \eqref{general-solution}, and in the final approximation we used the constraint $X>X_{c}$ from \eqref{bound-Xn} and the condition $\b/n\ll1$. Note that $S_{R}<2S_{BH}$ by the first equality in Eq. \eqref{EEc}, thus we have the approximate result $S_{R}\simeq 2S_{BH}$ and the Page time \eqref{pagetime} still holds for general case $\a < \b \ll 1$.

Now let us turn to the scrambling time. With the presence of the island, it was argued that the scrambling time $t_{\text{scr}}$ $( \text{with}\ t_{\text{scr}}\simeq \frac{1}{2\pi T_{h}}\log S_{BH}$ \cite{PHJP,YSLS}) can be estimated by the traveling time $\Delta t = t_a - t_0$ from $\rho_b$ to $\rho_a$, by assuming that a lightlike message sent from $\rho_b$ at $t_0$ to the island can
be instantly reconstructed from the Hawking radiation once it reaches the boundary of island $\rho_a$ at $t_{a}=t_{b}$. This was supported by the approximate result $\Delta t \simeq \frac{1}{2\pi T_{h}}\log S_{BH}$ obtained in Ref. \cite{Hashimoto:2020cas} under the condition $\tilde{\a}\ll\tilde{\b}\ll1$. However, generally we do \textit{not} obtain this approximate result for more general case $\a<\b\ll1$. To show this, let us calculate the traveling time, i.e.
\ba\label{scrambling-time}
\Delta t = t_a - t_0 = r_{*}(\rho_b)-r_{*}(\rho_a)
\simeq \frac{r_{h}}{n^2}\left(\b^2-\a^2+ 2 n\log\frac{\b}{\a}\right)
\simeq\frac{2r_{h}}{n}\log\frac{\b}{\a},
\ea
where we used \eqref{tortoise} and omitted the higher order terms of $\a,\b$.

For the special case $\a\ll\b/n\ll1$, the constraint $X\gg1$ should be considered, which equivalently requires that
\ba \label{constraint-limit}
\frac{2^{n}}{c \kappa_{D}(n+3) n^{n+2}} \cdot\frac{r_{h}^{n+1}}{{l_{p}}^{n+1}}\gg \frac{1}{\b^{n+3}}\gg1.
\ea
By plugging the solution \eqref{qes-X} into \eqref{scrambling-time}, we obtain
\ba\label{scr-time}
\Delta t
&=& \frac{2r_{h}}{n}\log\left[(n+3)X\right]= \frac{2r_{h}}{n}\log \frac{2^{n}\b^{n+3} \left(\frac{r_{h}}{l_{p}}\right)^{n+1}}{c \kappa_{D} n^{n+2}} \no
&=& \frac{2r_{h}}{n}\left[\log \frac{\Omega_{n+1}r_{h}^{n+1}}{4l_{p}^{n+1}} + \log\frac{2^{n+2}}{c \kappa_{D} n^{n+2}\Omega_{n+1}} - \log \frac{1}{\b^{n+3}} \right] \no
&\simeq& \frac{1}{2\pi T_{h}}\log {S_{BH}} \simeq t_{\text{scr}},
\ea
where in the first approximation we ignored the sub-leading term of $\b$ by using \eqref{constraint-limit} and also dropped the finite terms of $n, c$ (for finite $n, c$). This happens to give us the result of the scrambling time at leading order in special case $\a\ll\b/n\ll1$, similarly obtained in Ref. \cite{Hashimoto:2020cas} for $\tilde{\a}\ll\tilde{\b}\ll1$.

While for general case $\a<\b \ll1$, we have no analytical expression of $\Delta t$ without general analytical island solution. But we note that the island solution $\a/\b=x_{1}$ is monotonically increasing when $\b$ is decreasing for fixed $n, c$ and $ r_{h}$ (see the Fig. \ref{island-solution}). It means the traveling time \eqref{scrambling-time} will monotonically decrease when $\b$ decreases. If $\b\rightarrow \b_{\text{min}}$, we will have $\a/\b\rightarrow 1/(n+3)$, meanwhile
\ba
\Delta t \rightarrow (\Delta t)_{\text{min}}=\frac{2r_{h}}{n}\log (n+3)\sim \mathcal{O}(r_{h}),
\ea
which becomes obviously smaller than the scrambling time of order $\mathcal{O}(r_{h}\log (r_{h}/l_{p}))$ for black holes satisfying $r_{h}\gg l_{p}$. Thus for general case $\a<\b \ll1$, we do not receive the result like the scrambling time, thus $\Delta t\neq t_{\text{scr}}$. $\Delta t$ is highly dependent on the position of the inner boundary of the radiation region, where the $\b$-dependent terms can no longer be ignored for general $\a<\b \ll1$. Similar position dependent results of $\Delta t$ were also obtained in Refs. \cite{Hartman:2020swn,Gautason:2020tmk}, although the authors considered the case in which the inner boundary of the radiation region was far from the horizon of CGHS \cite{CGHS} black holes.

\section{ Conclusion and discussion }\label{Conclusion-discussion}
In this paper, we redefine a proper near-horizon condition in higher dimensions, which avoids the ill behavior of short-distance approximation in large dimensions under the near-horizon condition adopted in Ref. \cite{Hashimoto:2020cas}. The general island solution for $\a<\b\ll1$ is investigated in any $D\geq4$ dimension. An analytical island solution \eqref{qes} is obtained by assuming $\a\ll\b/n\ll1$, accompanying with the constraint \eqref{bound0}. And the existence of the island solution for general condition $\a<\b\ll1$ is also confirmed, meanwhile a more general constraint \eqref{bound-Xn} is obtained by the existence of the island solution in this case. We discuss the constraint \eqref{bound-Xn} for situation $\b\ll1$ from three equivalent aspects: constraint \eqref{bound1} on the size of the black holes, constraint \eqref{bound2-r} on the position of the inner boundary of the radiation region, and constraint \eqref{Semibound} on the value of $c\cdot\tilde{G}_{N}$. Moreover, the large dimension behavior of constraint \eqref{large-bound1} on the size of black hole is in agreement with the constraint in Ref. \cite{FDN}. These results are simply required by the presence of the island in situation $\b\ll1$, which can be understood as the unitary constraints in semiclassical gravity. The entanglement entropy of the radiation follows the Page curve and the Page time is obtained for $\a<\b\ll1$. While some doubts about the estimation of the scrambling time by $\Delta t$ are raised in this case. For special case $\a\ll\b/n\ll1$, $\Delta t$ gives the result as the scrambling time at leading order, i.e. $\Delta t\simeq t_{\text{scr}}$, as similarly obtained in Ref. \cite{Hashimoto:2020cas} for $\tilde{\a}\ll\tilde{\b}\ll1$. However, for general $\a<\b\ll1$, we find $\Delta t\neq t_{\text{scr}}$ generally, and $\Delta t$ is highly dependent on position of the inner boundary of the selected radiation region.

The existence of the constraint \eqref{bound-Xn} implied by the island solution for $\a<\b\ll1$ is surprising, which reveals some aspects from the unitarity in semiclassical gravity once we admit the island formalism. Some implications are shown in the paper, but much remains to be further investigated. We expect similar results can be obtained for other types of black holes, such as dynamical Schwarzschild black holes and the Reissner-Nordstr\"{o}m black holes and so on. In this paper, we only consider the situation where the inner boundary of the radiation region is near the horizon, i.e. $\b\ll1$, and the constraint \eqref{bound-Xn} is only valid for this situation. It is possible that there exists a more general and completed constraint for any $\b>0$ by requiring the existence of the island solution. However, the calculation of the entanglement entropy in general $D\geq4$ dimensional Schwarzschild black hole spacetime for any $\b>0$ is still a hard task. One may also consider another special situation where the inner boundary of radiation region is chosen to be far from the horizon, i.e. $\b \gg 1$, which is not considered in our paper. As the results of Refs. \cite{Arefeva:2021kfx,Arefeva:2022guf,Arefeva:2022cam,Ageev:2022hqc,He:2021mst} strongly indicate that there may also be a unitary constraint required by the existence of the island solution for situation $\b\gg 1$. Moreover, when ${r_{h}}/{l_{p}}\sim (c \cdot\chi)^{\frac{1}{n+1}}$ (or equivalently ${cl_{p}^{n+1}}/{r_{h}^{n+1}}\sim1/\chi$), we find no consistent island solution for situation $\b\ll1$, as we think it has been out of the semiclassical validity. This even occurs for a large black hole with size $r_{h}\sim n^{{3}/{2}}l_{p}\gg l_{p}$ in large dimension limit. The studies of large dimension gravity may help provide some insights into the future studies of quantum-gravity effects.

\section*{ Acknowledgements }
This work is supported by the National Natural Science Foundation of China under Grants No. 11675272 (J.R.S.) and No. 11975116 (F.W.S.), and the Jiangxi Science Foundation for Distinguished Young Scientists under Grant No. 20192BCB23007 (F.W.S.). W.C.G. is supported by Baylor University through the Baylor Physics graduate program.

\addcontentsline{toc}{section}{Appendices}
\addtocontents{toc}{\protect\setcounter{tocdepth}{0}}
\appendix
\section{ Short-distance approximation under original and new near-horizon condition }\label{Appendix A}
The island solution in higher dimensional spacetime has been studied in Ref. \cite{Hashimoto:2020cas} when inner boundary of the radiation region was chosen to be near the horizon, and the boundary of the island was assumed to be outside and near the horizon. In which the near-horizon condition was defined by $\tilde{\a}< \tilde{\b} \ll 1$ with parameters $\tilde{\a} \equiv \sqrt{{(r_{a} -r_{h})}/{r_h}}>0$ and $\tilde{\b} \equiv \sqrt{{(r_{b} -r_{h})}/{r_{h}}}>0$, thus $\tilde{\b}-\tilde{\a}< \tilde{\b} \ll 1$. However, we would like to point out that this near-horizon condition is not sufficient if one wants to simultaneously take a well-behaved short-distance approximation in large dimension case when applying the formula \eqref{Sm2}.

In any high dimension $D=n+3\geq4$, $n$ can also be treated as a variable and we need be careful when taking short-distance approximation under the original near-horizon condition $\tilde{\b}-\tilde{\a}< \tilde{\b} \ll 1$. To see the subtleties, we look at the geodesic distance and take the expansions for small $\tilde{\a},\tilde{ \b}$ ($\tilde{\a}$ is at most the order of $\tilde{\b}$), thus
\ba  \label{Lab1}
L &=& \int_{r_{a}}^{r_{b}} \frac{dr}{\sqrt{1- \left(\frac{r_{h}}{r} \right)^n}} \no
  &=& \frac{r_{h}}{\sqrt{n}}\left[\ 2(\tilde{\b}-\tilde{\a})+\frac{1}{6}(n+1)(\tilde{\b}^3-\tilde{\a}^3)\right.\no
  && \left. + \frac{1}{240}(n+1)(n-7)(\tilde{\b}^5-\tilde{\a}^5) +\ ... \ \right] \no
  &=& \frac{2r_{h}(\tilde{\b}-\tilde{\a})}{\sqrt{n}}\left( 1+ \mathcal{O}(n\tilde{\b}^{2})\right),
\ea
where $``..."$ are the higher order terms, and in the last equality we used $n\gg1$ to specify the order in large dimensions. Note that when $n$ is very large, a tighter condition $\tilde{\b}\ll {1}/{\sqrt{n}}$ (or $ n\tilde{\b}^2\ll1$) is required in order to effectively approximate $L$ by the first order term of $\tilde{\a},\tilde{ \b}$. This requires a much stronger constraint than the condition $\tilde{\b} \ll 1$ if $n\rightarrow\infty$. The short-distance approximation used in Ref. \cite{Hashimoto:2020cas}, i.e. $L\simeq 2r_{h}(\tilde{\b}-\tilde{\a})/\sqrt{n}$, only behaves well when the dimension is far below a critical dimension $n_{c}= {1}/{\tilde{\b}^2}$ with a small but finite $\tilde{\b}$. And huge corrections of order $\mathcal{O}(n\tilde{\b}^{2})$ would be required once $n \sim n_{c}$, even not taking $n\rightarrow\infty$. This tension between the original near-horizon condition and the short-distance approximation in large dimension case is always existing if we take a small but finite $\tilde{\b}$. This short-distance approximation in terms of small but finite $\tilde{\a},\tilde{ \b}$ is highly constrained by the dimension, and we need to find a more proper near-horizon condition which gets rid of this large dimension constraint.

The above problem can be avoided by using the coordinate $\ro$. Defining $\a\equiv\s{\ro_a-1}>0$ and $  \b\equiv\s{\ro_b-1}>0$, where $\ro_a={(r_{a}/r_{h})}^n$ and $\ro_b={(r_{b}/r_{h})}^n$. Also we assume boundary of the island is outside and near the horizon. And we take $\a< \b \ll 1$ as the new near-horizon condition with horizon at $\ro=1$, thus $\b-\a< \b \ll 1$. Considering the geodesic distance in coordinate $\ro$ and taking the expansions for small $\a,\b$ ($\a$ is at most the order of $\b$), i.e.
\ba  \label{Lab2}
L &=&  \int_{r_{a}}^{r_{b}} \frac{dr}{\sqrt{1- \left(\frac{r_{h}}{r} \right)^n}}=\int_{\ro_{a}}^{\ro_{b}} \frac{r_{h} \ro^{\frac{1}{n}} d\ro }{n\sqrt{\ro(\ro-1)}} \no
  &=& \frac{r_{h}}{n^4}\left[  2n^3(\b-\a)-\frac{1}{3}n^2(n-2)(\b^3-\a^3)\right.\no
  && \left. + \frac{1}{20}n(n-2)(3n-2)(\b^5-\a^5) +\ ...\  \right] \no
  &=& \frac{2r_{h}(\b-\a)}{n} \left( 1+ {\cal O}(\b^2) \right),
\ea
where $``..."$ represent the higher order terms, and in the last equality we used $n\gg1$ to specify the order in large dimensions. One can see the condition $\b \ll 1$ is sufficient to ensure the validity of taking the first-order approximation of small but finite $\a,\b$, even in large dimension limit $n\rightarrow\infty$. Since $\rho^{\frac{1}{n}}\rightarrow 1$ as $n\rightarrow\infty$ for finite $\rho$, which guarantees that each term in \eqref{Lab2} has the same order of $n$. Therefore, this new near-horizon condition can help us remove the tension with the short-distance approximation in large dimension case. So that we are able to estimate $L$ by the first order term of $\a, \b$ in any higher dimension. By definition we have $\tilde{\b}=\frac{\b}{\sqrt{n}}(1+{\cal O}(\b^2))$ (where we still use $n\gg1$ to specify the order in large dimensions) for $\b\ll1$, it turns out that the condition $\b\ll1$ is equal to the condition $\tilde{\b}\ll {1}/{\sqrt{n}}$ for finite $n$. But in order to properly reproduce the large dimension behavior, it's more proper to adopt condition $\b\ll1$ where we take a small but finite $\b$ (thus $ n \tilde{\b}^2\simeq\b^2\ll1$ is always satisfied), which avoids the problem when we take a small but finite $\tilde{\b}$ in large dimension $n \sim n_{c}$ (as $ n\tilde{\b}^2\sim1$, for which the higher order correction can not be ignored). Then it allows us to omit the higher order terms in \eqref{Lab1} or \eqref{Lab2} in any high dimension. Back to Section \eqref{With-island}.


\appendix

\end{document}